\documentclass[11pt,prc,twocolumn,nofootinbib,preprintnumbers,tightenlines,superscriptaddress]{revtex4-1}
\usepackage{amssymb,amsmath,amsfonts}
\usepackage{xcolor}
\usepackage{graphicx}
\usepackage{bm}
\usepackage{tabularx}
\usepackage[colorlinks=true]{hyperref}

\DeclareMathOperator{\re}{Re}
\DeclareMathOperator{\im}{Im}

\newcommand{\hM}{\hphantom{-}}
\newcommand{\dd}{\mathrm{d}}
\newcolumntype{Y}{>{\centering\arraybackslash}X}
\allowdisplaybreaks[1]
\makeatletter
\g@addto@macro\bfseries{\boldmath}
\makeatother

\begin{document}
\preprint{MITP-20-059}

\author{Bijaya Acharya}
\author{Vadim Lensky}
\author{Sonia Bacca}
\author{Mikhail Gorchtein}
\author{Marc Vanderhaeghen}
\affiliation{Institut f\"ur Kernphysik \&  Cluster of Excellence PRISMA$^+$,
 Johannes Gutenberg-Universit\"at  Mainz,  D-55128 Mainz, Germany}

\title{Dispersive evaluation of the Lamb shift in muonic deuterium \\ from chiral effective field theory}
%\title{Using  chiral effective field theory to inform dispersive evaluations of the Lamb shift in muonic deuterium \\}

\begin{abstract}
We  merge  the dispersive relation approach and the \emph{ab initio} method  to compute nuclear structure corrections to the Lamb shift in muonic deuterium.
We calculate the deuteron response functions and corresponding uncertainties up to next-to-next-to-next-to-leading order in chiral effective field theory and compare our results to selected electromagnetic data to test the validity of the theory. We then feed response functions calculated over a wide range of kinematics to the  dispersion-theory formalism and show that an improved accuracy is obtained compared to that with the use of available  experimental data in the dispersive analysis. This opens up the possibility of applying this hybrid method to other light muonic atoms and supplementing experimental data with \emph{ab initio} theory for kinematics where data are scarce or difficult to measure with the goal of reducing uncertainties in estimates of nuclear structure effects in atomic spectroscopy.

\end{abstract}

\date{\today}

\maketitle
%\tableofcontents
\section{Introduction}

In the past 10 years, Lamb shift measurements on muonic atoms have enabled the extraction of the charge radius of the proton and the deuteron with unprecedented precision. Inconsistencies with measurements performed on electronic systems have emerged, leading to the  so--called ``proton--radius puzzle'' ~\cite{Pohl:2010zza}  and the subsequent ``deuteron--radius puzzle''~\cite{Pohl669}. At least for the proton,  recent experimental results~\cite{Xiong:2019umf,Bezginov1007} seem to indicate that these differences are likely due to an underestimation of experimental uncertainties and systematics.

 Key to the success of muonic atom experiments is a precise knowledge of nuclear structure corrections due to the two-photon exchange diagram. The calculation of two-photon exchange in light muonic atoms 
 has attracted the interest of various theorists, who have devised different methods  to approach the problem (see, e.g., Ref.~\cite{Ji:2018ozm} and references therein).
 Nuclear structure uncertainties are typically at least one order of magnitude larger than the experimental precision, so that  any idea to reduce  uncertainties substantially helps the experiments to exploit their full potential.

The nuclear polarizability correction to the Lamb shift of muonic deuterium can be calculated using nucleon-nucleon (NN) interaction models. An assessment of model uncertainties in such theoretical calculations is of paramount importance when comparing with experiments. Based on dimensional analysis, Friar~\cite{Friar:2013rha} showed by an analysis in the zero-range approximation that it might be possible to reduce the uncertainty to about 1$\%$. Using the well-calibrated phenomenological AV18 NN potential~\cite{Wiringa:1994wb}, Pachucki performed a precision calculation of the polarizability correction~\cite{Pachucki:2011xr}, arriving at an $\mathcal{O}(1\%)$ estimate, obtained mainly from atomic physics considerations. This calculation was further improved in Ref.~\cite{Pachucki:2015uga} by including mass dependencies beyond the non-relativistic dipole form as well as the neutron polarizability contribution. A recent study based on analytic derivations of the longitudinal structure function in pionless effective field theory achieved a precision of about 5$\%$~\cite{Emmons:2020aov}. 

There have recently been two notable advancements in obtaining model-independent results with rigorous estimates of uncertainties for the polarizability correction to the Lamb shift of $\mu D$. Hernandez et al.~\cite{Hernandez:2014pwa,Hernandez:2017mof,Hernandez:2019zcm} have applied chiral effective field theory ($\chi$EFT) to perform rigorous analyses of statistical and systematic nuclear-structure uncertainties in theoretical calculations of the $\mu D$ Lamb shift. Carlson et al.~\cite{Carlson:2013xea} have performed a fully relativistic dispersive treatment of the two-photon-exchange (TPE) corrections to the $\mu D$ Lamb shift in which the TPE amplitude is given exactly by the physical elastic and inelastic $e D$ scattering. However, this fully data-driven approach yielded the rather large uncertainty of about 35\%, mainly due to the scarcity of data in the most relevant low-energy quasielastic kinematic regime. 

In this work, we build on these two recent achievements by using $\chi$EFT to generate the $e D$ scattering data that inform the dispersion-relations (DR) analysis to obtain model-independent results for the $\mu D$ Lamb shift using only controlled approximations. 
 It also allows us to validate the accuracy of the $\chi$EFT approach through a comparison with structure function data in the region where they are available. $\chi$EFT is a low-energy EFT of quantum chromodynamics that provides a description of hadronic and nuclear phenomena in terms of nucleons and pions as effective degrees of freedom. The nuclear Hamiltonian as well as couplings to external electroweak sources are systematically constructed as perturbative expansions in $k/\Lambda_b$ where $k\sim m_\pi$ is the typical (low) momentum scale of the process under study and $\Lambda_b$ is the scale beyond which the EFT breaks down. In a properly formulated EFT, one expects the calculated observables to inherit this $k/\Lambda_b$ expansion, which enables us to perform an order-by-order calculation and use the anticipated size of the ignored higher order terms as a measure of the theoretical uncertainty of the calculation.  

Here, we estimate the uncertainty due to higher-order terms in the $\chi$EFT expansion  as suggested in Ref.~\cite{Epelbaum:2014sza} using results obtained in an order-by-order expansion. For a given observable $\mathcal{O}$, the nuclear physics uncertainty is calculated as: 
\begin{align}
\label{band}
\Delta(\mathcal{O})\! & \!=\! {\rm max}\! \left\{
\chi^4 \left|\mathcal{O}^{(k^0)}\right|,
 \chi^3 \left|\mathcal{O}^{(k^1)}-\mathcal{O}^{(k^0)}\right|, \right. \\
\nonumber
& \left.
 \chi^2 \left|\mathcal{O}^{(k^2)}-\mathcal{O}^{(k^1)}\right|, \chi \left|\mathcal{O}^{(k^3)}-\mathcal{O}^{(k^2)}\right|\right\}\,,
\end{align}
 where $\chi=k/\Lambda_b=m_\pi/\Lambda_b$ is the estimate for the chiral expansion parameter, and the superscripts indicate the order of the $\chi$EFT calculation, for which we consider corrections to both currents and interactions appearing at their respective orders.

\section{Evaluation of the Response Functions in $\chi$EFT}

The cross section for the disintegration of an unpolarized deuteron by an electron of energy $\epsilon$ in the rest frame of the deuteron can be expressed in the one-photon-exchange approximation as
\begin{eqnarray}
\label{eq:crosssec}
\frac{\mathrm{d}^2\sigma}{\mathrm{d}\Omega\,\mathrm{d}\epsilon^\prime}
&= & \sigma_\mathrm{Mott} \left[ \frac{(\nu^2-\mathbf{q}^2)^2}{\mathbf{q}^4} \, R_L(\nu,q) \right. \\
\nonumber
 && - \left. \left(\frac{\nu^2-\mathbf{q}^2}{2\mathbf{q}^2} -\tan^2\frac{\theta}{2} \right) \, R_T(\nu,q) \right] \,.
\end{eqnarray}
Here, $\sigma_{\rm Mott}$ is the Mott cross section, given by 
\begin{equation}
 \sigma_M = \frac{\alpha_\mathrm{em}^2}{ \, 4\epsilon^2} \,
 \frac{\cos^2\frac{\theta}{2}}{\sin^4\frac{\theta}{2}}\,,
\end{equation}
and
\begin{equation}
 \nu=\epsilon-\epsilon^\prime\,,
\end{equation} 
where $\epsilon^\prime$ is the energy of the scattered electron in the laboratory frame at an angle $\theta$,
with 
\begin{eqnarray}
\mathbf{q}^2&=&\epsilon^2+{\epsilon^\prime}^2 -2\,\sqrt{\epsilon^2-m_e^2}\times  \\
\nonumber
&&\sqrt{{\epsilon^\prime}^2-m_e^2}\,\cos\theta-2m_e^2\,,
\end{eqnarray}
$\mathbf{q}$ being the three-momentum transfer, $q=\vert\mathbf{q}\vert$.
 $R_L$ and $R_T$ are the inelastic response functions of the electric charge and the transverse electromagnetic current operators, defined as
\begin{widetext} 
\begin{equation}
\label{eq:r_l}
  R_L(\nu,q) = \frac{1}{3} \sum_{s_d} \sum_{S^\prime S_z^\prime} \sum_{T^\prime} \int\frac{\mathrm{d}^3p}{(2\pi)^3} 
                  \,\frac{1}{2} \, \delta(\nu+M_d-E_+-E_-)
                  \vert \langle\psi_{\mathbf{p},S^\prime S_z^\prime,T^\prime 0}\vert \rho \vert\psi_{d,s_d}\rangle \vert^2
\end{equation}
and 
\begin{equation}
\label{eq:r_t}
  R_T(\nu,q) = \frac{1}{3} \sum_{s_d} \sum_{S^\prime S_z^\prime} \sum_{T^\prime} \int\frac{\mathrm{d}^3p}{(2\pi)^3} 
                  \,\frac{1}{2} \, \delta(\nu+M_d-E_+-E_-)
                  \sum_{\lambda=\pm1} \vert  \langle\psi_{\mathbf{p},S^\prime S_z^\prime,T^\prime 0}\vert j_\lambda\vert\psi_{d,s_d}\rangle \vert^2\,.
\end{equation}
\end{widetext}
Here, $\rho=j^0$ is the charge operator and $j_\lambda$ are the spherical components of the three-vector current operator ${\bf j}$ at four-momentum transfer $(\nu,\mathbf{q})$. The initial state is the deuteron ground state, denoted $| \psi_{d,s_d}\rangle$,  where $s_d$ is the projection of the total angular momentum, whereas 
the final state, denoted $|\psi_{\mathbf{p},S^\prime S_z^\prime,T^\prime T_z^\prime}\rangle $, is the $pn$ scattering state with the relative momentum, total isospin, isospin projection, total spin and spin projection given by $\mathbf{p}$, $T^\prime$, $T_z^\prime$, $S^\prime$ and  $S_z^\prime$ respectively. In the rest frame of the deuteron, the energies of the final-state nucleons are $E_\pm = \sqrt{(\mathbf{q}/2 \pm \mathbf{p})^2+m_N^2}$. Throughout this paper, we use $m_N$ to denote the isospin-averaged nucleon mass in the relevant NN isospin channel. The deuteron mass is denoted $M_d$. 

The cross section in Eq.~\eqref{eq:crosssec} can equivalently be expressed as 
\begin{eqnarray}
&\frac{\mathrm{d}^2\sigma}{\mathrm{d}\Omega\,\mathrm{d}\epsilon^\prime}
&= \sigma_\mathrm{Mott} \times \\
\nonumber
 &&\left[ \frac{2}{M_d}\tan^2\frac{\theta}{2}\, F_1(\nu,Q^2)
+\frac{1}{\nu}F_2(\nu,Q^2)\right]\,,
\end{eqnarray}
in terms of the unpolarized response functions $F_1$ and $F_2$ which are better suited for a dispersive calculation, as explained in Sec.~\ref{sec:DR_formalism}. These are related to the response functions $R_L$ and $R_T$, which were defined in terms of the Coulomb and transverse operators, by 
\begin{align}
    F_1(\nu,Q^2) & = \frac{M_d}{2} R_T(\nu, q)\,, \\
    F_2(\nu,Q^2) & =  \frac{\nu}{2}\frac{Q^2}{Q^2+\nu^2}\bigg[R_T(\nu,q)\\
\nonumber
                &  + \frac{2Q^2}{Q^2+\nu^2}R_L(\nu,q)\bigg]\,,
\end{align}
where $Q^2=\mathbf{q}^2-\nu^2$ is the virtuality of the exchanged photon.

The nuclear wave functions in Eqs.~\eqref{eq:r_l} and \eqref{eq:r_t} are obtained by solving the Lippmann-Schwinger equation for the NN system with the $\chi$EFT interactions in Ref.~\cite{PhysRevC.96.024004}. The electromagnetic current operators we use were first derived in $\chi$EFT in Refs.~\cite{Park:1995pn,Pastore:2008ui,Kolling:2009iq}. They can be expressed as sums of one-body (1B) and two-body (2B) operators that act on nucleonic degrees of freedom. In principle, in a consistent $\chi$EFT calculation, the nucleon form factors that feature in these current operators should be calculated to the same order in $\chi$EFT at which the operators are truncated. However, the convergence of these nucleon-structure corrections is rather slow~\cite{Kubis:2000zd}. It has therefore become a common practice to use phenomenological form factors to represent the sum of the nucleon structure diagrams, which makes the calculations of nuclear systems less sensitive to inaccuracies in the single-nucleon sector~\cite{Phillips:2016mov}.  If not otherwise explicitly indicated, we use the recent parametrization from Ref.~\cite{Tomalak} for the proton and neutron form factors $G_{E/M}^{p/n}(Q^2)$, which incorporates the small proton radius obtained from the muonic hydrogen measurement~\cite{Pohl:2010zza}.

As in Ref.~\cite{Bijaya_Neutrino}, we perform a $\chi$EFT expansion of the electromagnetic operators $\rho$ and ${\bf j}$ up to  order $k^3$, which corresponds to next-to-next-to-leading order (N2LO) in the interactions where there are no contributions at order $k^1$.
The charge operator $\rho$ first appears at the leading ($k^0$) order and is purely 1B up to the $\chi$EFT order we consider in this work. It is given by the sum of isoscalar and isovector components, $\rho=\rho^S+\rho^V$. The matrix elements of these operators between eigenstates of two-nucleon relative momenta can be written as  
\begin{widetext}
\begin{equation}
 \langle\mathbf{p^\prime}\vert\rho^{S/V}(q^\mu)\vert\mathbf{p}\rangle = G_E^{S/V}(Q^2) \dfrac{1}{\sqrt{1+\frac{Q^2}{4m_N^2}}} \, \tau_1^{S/V}\,\delta^{(3)}\left(\mathbf{p^\prime}-\mathbf{p}-\frac{\mathbf{q}}{2}\right)+(1\leftrightarrow2)\,.
\end{equation}
For the isoscalar case, the isospin operator $\tau_n^S$ is one-half times the identity operator whereas the isovector isospin operator $\tau_n^V$ is $\tau_{n,z}/2$. The electric form factors $G_E^{S/V}$ are related to the proton and neutron electric form factors by $G_E^{S/V}(Q^2)=G_E^p(Q^2) \pm G_E^n(Q^2)$. 

The 1B current operator ${\bf j}$ first contributes at order $k^1$. It consists of the so-called convection and spin-magnetization currents,
\begin{align}
 \langle\mathbf{p^\prime}\vert\mathbf{j}_\mathrm{1B}^{S/V}(q^\mu)\vert\mathbf{p}\rangle =  \bigg( G_E^{S/V}(Q^2)\frac{\mathbf{p^\prime}+\mathbf{p}}{2m_N} 
   - i \, G_M^{S/V}(Q^2)\frac{\mathbf{q}\times\bm{\sigma}_1}{2m_N}\bigg)
    \, \tau_1^{S/V} \,\delta^{(3)}\left(\mathbf{p^\prime}-\mathbf{p}-\frac{\mathbf{q}}{2}\right)+(1\leftrightarrow2)\,.
\end{align}

The 2B current operators are purely isovector up to order $k^3$, which is the highest order we consider here. They are given by the sum of the so-called seagull and pion-in-flight diagrams which can be written as 
\begin{align}
\langle\mathbf{p^\prime}\vert\mathbf{j}_\mathrm{2B}(q^\mu)\vert\mathbf{p}\rangle = -i \,   \frac{g_A^2}{4f_\pi^2}  
\bigg( \bm{\sigma}_1-\mathbf{k}_1\frac{\bm{\sigma}_1\cdot\mathbf{k}_1}{m_\pi^2+\mathbf{k}_1^2}\bigg) \,\frac{\bm{\sigma}_2\cdot\mathbf{k}_2}{m_\pi^2+\mathbf{k}_2^2}
 \, \left(\bm{\tau}_1\times\bm{\tau}_2\right)_z \,\delta^{(3)}(\mathbf{k}_1+\mathbf{k}_2-\mathbf{q})+ (1\leftrightarrow 2)\,,
\end{align}
where $\mathbf{k}_{1,2}=\mathbf{p}^\prime_{1,2}-\mathbf{p}_{1,2}$, $f_\pi=92.3$~MeV is the pion-decay constant and $g_A=1.2723$ is the axial coupling constant~\cite{PDG2020}.
\end{widetext}

\section{Comparison  to electromagnetic data}

Before using $\chi$EFT response functions for evaluation of the Lamb shift in the dispersion formalism, we compare our results  obtained using NN interactions from Ref.~\cite{PhysRevC.96.024004} with deuteron photo- and electro-dissociation experiments. To this end, we first calculate the photodissociation cross section, 
\begin{equation}
\sigma_{\gamma d}(\nu) = \frac{2\pi^2}{\nu} \alpha  R_T(\nu,\nu)\,,
\end{equation}
both in the impulse approximation, which uses only one-body currents (1B), as well as with the one- and two-body currents explicitly included (1B+2B). We also perform a calculation in the so-called dipole approximation, where effects of 2B currents are implicitly included via the Siegert theorem~\cite{Siegert}.  
\begin{figure}[h]
\centering
    \includegraphics[width=\columnwidth]{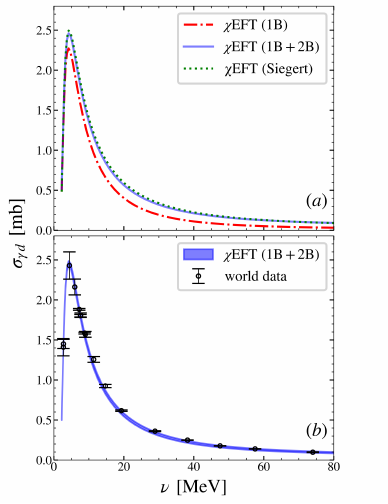}
    \caption{Deuteron photodissociation cross section. (a): 1B and 1B+2B  results in comparison to the Siegert operator. (b):  1B+2B  results with the corresponding theoretical uncertainty band compared against data from Refs.~\cite{PhysRevC.32.1825,PhysRevLett.57.1542}.}
    \label{fig:photo_diss}
\end{figure}

 Although current conservation is not strictly satisfied in a $\chi$EFT calculation due to the artifacts of regularization of the nuclear Hamiltonian, we obtain a very good agreement between the Siegert calculation and the one that explicitly includes one- and two-body currents.  Figure~\ref{fig:photo_diss}(a) clearly shows that we achieve current conservation to a very good approximation. In Fig.~\ref{fig:photo_diss}(b), we show the excellent agreement we obtain with the highly precise experimental photodissociation world data compiled from Refs.~\cite{PhysRevC.32.1825,PhysRevLett.57.1542} with the 1B+2B result. The latter includes a theory uncertainty band obtained using Eq.~\eqref{band} by varying the chiral orders consistently in the interactions as well as in the currents. Throughout this work, uncertainties are estimated for $\Lambda_b=500$~MeV, which corresponds to the cutoff used in the interactions. In Fig.~\ref{fig:photo_diss}, we used currents up to order $k^3$ and the N2LO potential, which also includes all contributions to the NN interaction up to order $k^3$. Such a consistent truncation of the expansions in current and interaction is a necessary condition for current conservation to be formally satisfied. Subsequent results, unless otherwise stated, are obtained with the N3LO potential, which is more widely used in the literature. Such calculations, which we call order $k^{3+}$, contain order $k^4$ effects in the NN interactions but not in the currents, and agree with the order $k^3$ results well within the uncertainties given by 
Eq.~\eqref{band}.

To provide another reference point, the photodissociation cross section and the transverse response function can be related to the dipole polarisabilities of the deuteron. The famous Baldin sum rule~\cite{Baldin:1960} connects $\sigma_{\gamma d}$ with the sum of the electric $\alpha_{E1}$ and magnetic $\beta_{M1}$ polarisabilities of the deuteron:
\begin{equation}
\frac{1}{2\pi^2}\int\limits_{\nu_\mathrm{thr}}^\infty\mathrm{d}\nu\,\frac{\sigma_{\gamma d}(\nu)}{\nu^2}\, = \alpha_{E1}+\beta_{M1}\,,
\label{eq:BaldinSR}
\end{equation}
whereas $\beta_{M1}$ is related to the slope of the transverse response function at $Q^2=0$ via~\cite{Gorchtein:2015eoa}
\begin{equation}
\beta_{M1} = \frac{2\alpha}{M_d}\int\limits_{\nu_\mathrm{thr}(0)}^{\nu_\mathrm{max}(0)}
\frac{\mathrm{d}\nu}{\nu} \frac{\mathrm{d}\hphantom{Q^2}}{\mathrm{d}Q^2}F_1(\nu,Q^2)\big|_{Q^2=0}\,.
\label{eq:betaSR}
\end{equation}
Equation~(\ref{eq:betaSR}) neglects the recoil corrections in the integration limits, as well as the contributions of the deuteron and nucleon charge radii. The neglected contributions turn out to be very small for the deuteron (at the level of $\lesssim 1\%$~\cite{Gorchtein:2015eoa}). Furthermore, since the theoretical result for $\sigma_{\gamma d}(\nu)$ only contains the photodissociation contribution (omitting the pion production and other higher-energy channels), the evaluation of Eq.~(\ref{eq:BaldinSR}) will
also neglect the nucleon contribution to the deuteron polarizabilities (which is also expected to be at the $\lesssim 1\%$ level).

We now present our numerical results starting with the Baldin sum rule. Our calculation predicts
\begin{equation}
\label{eq:BaldinSR_result}
\alpha_{E1}+\beta_{M1} = 0.698(18)\text{ fm}^3\,,
\end{equation}
which coincides with the empirical result~\cite{Friar:1983zza} $\alpha_{E1}+\beta_{M1}=0.69(0.04)\text{ fm}^3$, as one expects given the good description of the photodissociation cross section.

Together with the Baldin sum rule, the sum rule for the magnetic polarizability, Eq.~(\ref{eq:betaSR}), yields 
\begin{eqnarray}
\nonumber
 \alpha_{E1}&=& 0.626(18)   \text{ fm}^3, \\
 \beta_{M1} &=& 0.0715(15)   \text{ fm}^3\,.
\label{eq:alphabeta}
\end{eqnarray}
The quoted uncertainties both in the Baldin sum rule and in the values of $\alpha_{E1}$ and $\beta_{M1}$ are  estimates of higher-order terms in the $\chi$EFT expansion. 

The value of $\alpha_{E1}$ compares well with the empirical extraction of Ref.~\cite{Friar:1983zza}, $\alpha_{E1}=0.61(0.04)\text{ fm}^3$. It is also instructive to compare the dominant 
dipole contribution to $\alpha_{E1}$ calculated via the dipole sum rule~\cite{Friar:1984zzb,Friar:1997jy}:
\begin{equation}
\label{eq:dipoleSR}
\alpha_{E1}^\mathrm{dipole} = \frac{1}{2\pi^2}\int\limits_{\nu_\mathrm{thr}}^\infty\mathrm{d}\nu\,\frac{\sigma_{\gamma d}^\mathrm{dipole}(\nu)}{\nu^2}\,,
\end{equation}
where only the dipole operator contribution is retained in the cross section. Evaluating this sum rule, we obtain
\begin{equation}
\label{eq:dipoleSR_result}
\alpha_{E1}^\mathrm{dipole}=0.6340(10)\text{ fm}^3\,,
\end{equation}
in agreement with the result obtained averaging various phenomenological potentials~\cite{Friar:1997jy}, $\alpha_{E1}^\mathrm{dipole}=0.6328(17) \text{ fm}^3$.

In contrast to $\alpha_{E1}$, which is dominated by the long-range dipole contribution, the deuteron magnetic polarisability $\beta_{M1}$ is affected by shorter-range mechanisms through the magnetic transition, as pointed out in Ref.~\cite{Friar:1983zza}, whose model gives $\beta_{M1}=0.071(6) \text{ fm}^3$. Our prediction for $\beta_{M1}$ is in agreement with this, albeit with a significantly smaller uncertainty estimate.

%\begin{align}
%\sigma(\alpha_{E1})  = &\rm{max} \left\{
%\chi^4 \left|\alpha_{E1}^{(k^0)}\right|,\
%\chi^3 \left|\alpha_{E1}^{(k^1)}-\alpha_{E1}^{(k^0)}\right|,\ \right. \\
%\nonumber
%&\left. \chi^2 \left|\alpha_{E1}^{(k^2)}-\alpha_{E1}^{(k^1)}\right|,\ \chi \left|\alpha_{E1}^{(k^3)}-\alpha_{E1}^{(k^2)}\right|\right\}\,,
%\end{align}
%where the superscripts show the $\chi$EFT orders, and analogously for $\beta_{M1}$.

\begin{figure}[ht]
\centering
    \includegraphics[width=\columnwidth]{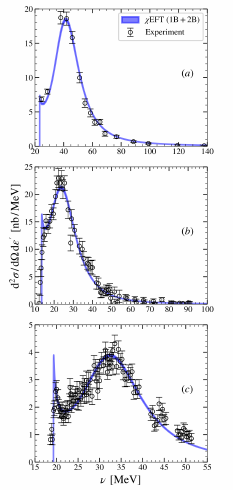}
    \caption{Electrodissociation cross section of the deuteron with data from $(a)$ Ref.~\cite{PhysRevC.38.800} at $\epsilon=292.8~\mathrm{MeV},\,\theta=60^\circ$,  $(b)$
    Ref.~\cite{PhysRev.116.1257} at $\epsilon=175~\mathrm{MeV},\,\vert\mathrm{q}\vert=206~\mathrm{MeV}$, and  $(c)$ Ref.~\cite{YEARIAN1964234} at $\epsilon=146.9~\mathrm{MeV},\,\theta=135^\circ$.}
    \label{fig:electron_scatter}
\end{figure}

We now turn our attention to the electrodissociation cross section of the deuteron calculated using Eq.~\eqref{eq:crosssec}. Figure~\ref{fig:electron_scatter} shows this cross section at various kinematics, along with experimental data from Refs.~\cite{PhysRevC.38.800,PhysRev.116.1257,YEARIAN1964234}.  An excellent agreement is obtained for both the final-state interaction effects near threshold and the quasielastic peak, with the theoretical precision surpassing the experimental uncertainty.

\begin{figure}[ht]
\centering
    \includegraphics[width=\columnwidth]{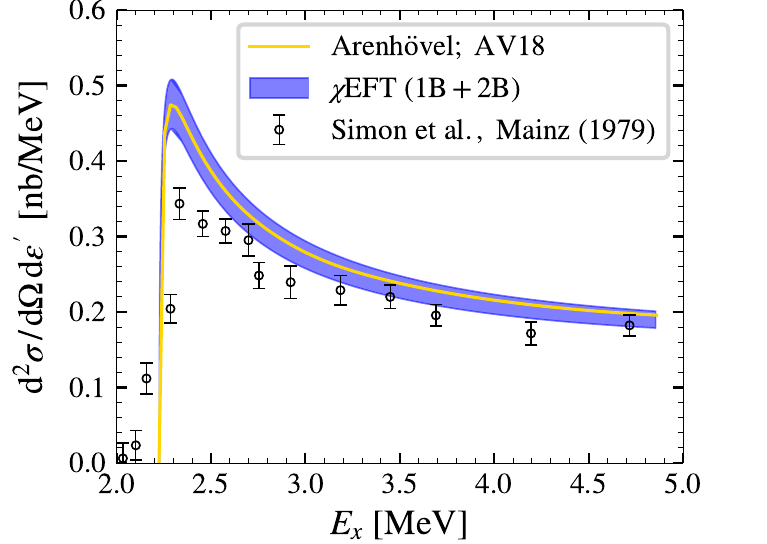}
  \caption{$\chi$EFT results (blue band) for the electrodissociation cross section of the deuteron along with theoretical calculation of Ref.~\cite{Arehoevel_priv_comm} (yellow line) and data from Ref.~\cite{SIMON1979277} at $\epsilon=222.6~\mathrm{MeV},\,\theta=157^\circ$. }
    \label{fig:near_threshold}
\end{figure}

 In Fig.~\ref{fig:near_threshold}, we show the electrodissociation cross section calculated at kinematics that correspond to low values of the excitation energy 
$E_x = \nu-\mathbf{q}^2/(2M_d)$. In these kinematics, the data suffer from detector resolution issues as evident from the leakage of data into the kinematically forbidden values, $E_x<B_d$, where $B_d=2.22446$~MeV is the binding energy of the deuteron~\cite{Mohr:2015ccw}. A closer comparison between theory and experiment would therefore require convolving the theoretical curves with the detector resolution, which has not been performed here. However, we do note the excellent agreement between our $\chi$EFT results and the theoretical calculation from Ref.~\cite{Arehoevel_priv_comm} performed with AV18 interactions and phenomenological one- and two-body currents.

\section{Dispersion Theory Formalism}
\label{sec:DR_formalism}
%\subsection{Two-photon exchange correction at  $O(\alpha^5)$}
We briefly review the relativistic DR formalism used in the evaluation of the
nuclear structure contribution to the TPE corrections (see Refs.~\cite{Carlson:2013xea,Carlson:2011zd,Gorchtein:2015eoa} for details). Our strategy is to then feed the response functions calculated in $\chi$EFT to the DR analysis.

\begin{widetext}
The nuclear structure TPE contribution to the Lamb shift at $O(\alpha^5)$ is related to the spin-independent
part of the forward double virtual Compton (VVCS) tensor,
\begin{align}
T^{\mu\nu} &= \frac{i}{8\pi M_d}\int \dd^4 x\, e^{i q x}\left\langle p\right|T\left(j^\mu(x) j^\nu(0)\right)\left|p\right\rangle \nonumber \\
& = \left(-g^{\mu\nu}+\frac{q^\mu q^\nu}{q^2}\right)T_1(\nu,Q^2)
+\frac{1}{M_d^2}\left(p^\mu-\frac{p\cdot q}{q^2}q^\mu\right)\left(p^\nu-\frac{p\cdot q}{q^2}q^\nu\right)
T_2(\nu,Q^2)\,.
\end{align}
Here, $q^2 = -Q^2$, $\nu = p\cdot q/M_d$, and $T_{1,2}(\nu,Q^2)$ are the two scalar VVCS amplitudes.
The dependence of the amplitude on the deuteron spin is trivial $\propto \varepsilon_d\cdot\varepsilon_{d'}^*$ and is omitted. The TPE contribution to the energy of the
$(n\ell)$ atomic level is given in terms of $T_{1,2}(\nu, Q^2)$ by
\begin{align}
\label{eq:lambshift_1}
    \Delta E_{n\ell}&=\frac{8\alpha^2 m}{\pi}\left[\phi_{n\ell}(0)\right]^2 \int\dd^4 Q
    \frac{(Q^2-2\nu^2)T_1(\nu,Q^2)-(Q^2+\nu^2)T_2(\nu,Q^2)}{Q^4(Q^4-4m^2\nu^2)}\,,
\end{align}
where $m$ is the lepton mass, $|\phi_{n\ell}(0)|^2= m_r^3\alpha^3/(\pi n^3)\delta_{\ell0}$ is the atomic wave function squared at the origin,  and $m_r=m M_d/(m+M_d)$ is the reduced mass.

\end{widetext}

The amplitudes $T_{1,2}(\nu,Q^2)$
are even functions of $\nu$, and their imaginary parts are connected by the optical theorem to the
unpolarized structure functions $F_{1,2}(\nu,Q^2)$ through
\begin{align}
    \im T_1(\nu,Q^2) & = \frac{1}{4M_d}F_1(\nu,Q^2)\,,\\
    \im T_2(\nu,Q^2) & = \frac{1}{4\nu}F_2(\nu,Q^2)\,.
\end{align}
It is convenient to separate the elastic contribution to the structure functions,
\begin{align}
F_1^\mathrm{el}(\nu,Q^2)  = & \frac{1}{3}(1+\tau_d)G_M^2(Q^2)\,\delta(1-x_d)\,,\\
F_2^\mathrm{el}(\nu,Q^2)  = & \left[G_C^2(Q^2)+\frac{2}{3}\tau_d G_M^2(Q^2) \right. \\
\nonumber
&+ \left. \frac{8}{9}\tau_d^2G_Q^2(Q^2)\right]\,\delta(1-x_d)\,,
\end{align}
where $G_C$, $G_M$, and $G_Q$ are the deuteron charge, magnetic, and quadrupole form factors, 
$\tau_d=Q^2/(4M_d^2)$, and $x_d = Q^2/(2M_d\,\nu)$. These contributions correspond to the pole
parts of the VVCS amplitudes,
\begin{align}
    T_{1,2}(\nu,Q^2) = T^\mathrm{pole}_{1,2}(\nu,Q^2) + T^\mathrm{non-pole}_{1,2}(\nu,Q^2)\,,
\end{align}
where the poles lie at $\nu=\pm Q^2/(2M_d)$ and are expressed through the deuteron form factors.
Upon subtracting the point-like deuteron and the finite-size effects, $T_{1,2}^\mathrm{pole}(\nu,Q^2)$
provide the elastic contribution to the energy shifts.

From this point onward we assume that the pole parts are subtracted from $T_{1,2}(\nu,Q^2)$, and, correspondingly,
that $F_{1,2}(\nu,Q^2)$ contain only the inelastic contributions.
Considering the known high-energy behavior of the structure functions~\cite{Drechsel:2002ar,Pasquini:2018wbl},
one obtains the DR espressions for $T_1(\nu,Q^2)$ and $T_2(\nu,Q^2)$---a once-subtracted one
and an unsubtracted one, respectively:
\begin{align}
\re T_1(\nu,Q^2) & = \bar{T}_1(0,Q^2)
 \nonumber \\
&  +\frac{\nu^2}{2\pi M_d} \int\limits_{\nu_\mathrm{thr}(Q^2)}^\infty
\frac{\mathrm{d}\omega}{\omega} 
\frac{F_1(\omega,Q^2)}{\omega^2-\nu^2}, \nonumber\\
\re T_2(\nu,Q^2) & =%\hphantom{\bar{T}_1(0,Q^2)+} 
\frac{1}{2\pi }\int\limits_{\nu_\mathrm{thr}(Q^2)}^\infty
\mathrm{d}\omega\, \frac{F_2(\omega,Q^2)}{\omega^2-\nu^2}\,,
\label{eq:DRs}
\end{align}
where $\nu_\mathrm{thr}(Q^2)=E_d+(E_d^2+Q^2)/(2M_d)$ is the inelastic threshold, and $\bar{T}_1(0,Q^2)$ 
is the subtraction function. The total TPE contribution is thus split into three parts---elastic, inelastic,
and subtraction. Using the DR in Eqs.~(\ref{eq:DRs}), as well as their equivalents
for the pole parts, one arrives at the following expressions for the three terms, with the  contributions of the point-like charge and charge radius of the deuteron
%and finite-size effects 
removed from the elastic part to avoid double-counting~\cite{Carlson:2013xea}:
\begin{widetext}
\begin{align}
\label{eq:contrib_elastic}
\Delta \bar{E}_{n\ell}^\mathrm{el} & = \frac{m \alpha^2}{M_d(M_d^2-m^2)}[\phi_{n\ell}(0)]^2
\int\limits_0^\infty\frac{\mathrm{d}Q^2}{Q^2} 
 \times\left\{
\frac{2}{3}G_M^2(Q^2)
(1+\tau_d)\hat{\gamma}_1(\tau_d,\tau_l)
%\left(\frac{\gamma_1(\tau_d)}{\sqrt\tau_d}-\frac{\gamma_1(\tau_l)}{\sqrt\tau_l}\right)
- \right. \nonumber\\
& \qquad \left. %\left(\frac{\gamma_2(\tau_d)}{\sqrt\tau_d}-\frac{\gamma_2(\tau_l)}{\sqrt\tau_l}\right)
\left[\frac{G_C^2(Q^2)-1}{\tau_d}+\frac{2}{3}G_M^2(Q^2)+\frac{8}{9}\tau_d G_Q^2(Q^2)\right]
\hat{\gamma}_2(\tau_d,\tau_l)
\right. +16M_d^2\frac{M_d-m}{Q}G_C'(0)
\bigg\}\,, \\
\label{eq:contrib_inelastic}
\Delta E_{n\ell}^\mathrm{inel}
&=-\frac{2\alpha^2}{M_d m}[\phi_{n\ell}(0)]^2
\int\limits_0^\infty\frac{\mathrm{d}Q^2}{Q^2}
\int\limits_{\nu_\mathrm{thr}(Q^2)}^\infty
\frac{\mathrm{d}\nu}{\nu}
\left[\tilde\gamma_1(\tau,\tau_l) F_1(\nu,Q^2)
+\frac{M_d\nu}{Q^2}\tilde\gamma_2(\tau,\tau_l)F_2(\nu,Q^2)\right]\,, \\
\label{eq:contrib_subtraction}
\Delta \bar{E}_{n\ell}^\mathrm{subt} &= \frac{4\pi\alpha^2}{m}[\phi_{n\ell}(0)]^2
\int\limits_0^\infty\frac{\mathrm{d}Q^2}{Q^2}
\frac{\gamma_1(\tau_l)}{\sqrt{\tau_l}}\left[\bar{T}_1(0,Q^2)-\bar{T}_1(0,0)\right]\,.
\end{align}

\end{widetext}
Here, we denote $\tau=\nu^2/Q^2$, $\tau_l=Q^2/(4m^2)$, and the weighting functions are defined as
\begin{align}
\gamma_1(x) & = (1-2x)\sqrt{1+x}+2x^{3/2}\,,\\
\gamma_2(x) & = (1+x)^{3/2}-x^{3/2} -\frac{3}{2}\sqrt{x}\,,\\
\hat{\gamma}_{1,2}(x,y) & = \frac{\gamma_{1,2}(x)}{\sqrt{x}}-\frac{\gamma_{1,2}(y)}{\sqrt{y}}\,,\\
\tilde{\gamma}_1(x,y) & = \frac{\sqrt{x}\,\gamma_1(x)-\sqrt{y}\,\gamma_1(y)}{x-y}\,,\\
\tilde{\gamma}_2(x,y) & = \frac{\hat{\gamma}_2(x,y)}{y-x}\,.
\end{align}
The expressions in Eqs.~\eqref{eq:contrib_elastic}-\eqref{eq:contrib_subtraction} follow from Eqs.~\eqref{eq:lambshift_1}-\eqref{eq:DRs} upon applying a Wick rotation $\nu = i q_0$, introducing  four-dimensional hyperspherical coordinates 
%with $q_0$ chosen as the polar axis and $Q$ as the radial distance. After that, an integration 
and finally integrating over the hyperspherical angles.
Note that the amplitude $\bar{T}_1(0,Q^2)$ in the subtraction term also gets an additional subtraction to
remove the effects of the pointlike deuteron (the Thomson term).

\subsection{Subtraction}
In general, the subtraction contribution presents a non-trivial problem since the general $Q^2$ dependence of the subtraction function cannot be directly related to observable quantities, and calculating this term may involve a significant deal of modeling (see the relevant discussion for the hydrogen Lamb shift in Refs.~\cite{Carlson:2011zd,Birse:2012eb}). Fortunately, the case of composite nuclei features two very different characteristic energy scales: nuclear, of the order of a few MeV, and hadronic, of the order of a (few) hundred MeV. They correspond to whether the photons probe the nuclear structure (starting from the first inelastic threshold) or the internal structure of the nucleons within the nucleus (starting from the pion production threshold).
Since typical nuclear cross sections fall off rather rapidly as the energy transfer increases past a certain point (of a typical nuclear scale), one can expect that there is a region of photon energies where the nuclear structure response is already very small, whereas the nucleon structure has not yet started to be probed. This is indeed universally seen in, e.g., nuclear photoabsorption data, and this manifest separation of scales allows one to write a sum rule for the subtraction amplitude that selects the nuclear contribution to the structure functions~\cite{Gorchtein:2015eoa}:
\begin{align}
\label{eq:nuclear_SR}
&    \bar{T}_1(0,Q^2) - \bar{T}_1(0,0)  = \qquad\qquad\qquad\qquad\\
\nonumber 
&\quad \frac{1}{2\pi M_d} \int\limits_{\nu_\mathrm{thr}(Q^2)}^{\nu_\mathrm{max}(Q^2)}
    \frac{\mathrm{d}\nu}{\nu}\left[F_1(\nu,Q^2)-F_1(\nu,0)\right]\,,
\end{align}
where $\nu_\mathrm{max}(Q^2)$ has to be chosen in the above region of intermediate energies where the photoabsorption cross sections are small. With the sum rule of Eq.~\eqref{eq:nuclear_SR}, the subtraction contribution to the Lamb shift can then be expressed in terms of the structure function $F_1$ as 
\begin{align}
\label{eq:contribution_subtraction_final}
    \Delta \bar{E}^\mathrm{subt}_{n\ell} & = \frac{8\alpha^2}{M_d}\left[\phi_{n\ell}(0)\right]^2 
    \int\limits_0^\infty \frac{\mathrm{d}Q}{Q^2}\,\gamma_1(\tau_l)\\
\nonumber
 &  \times   \int\limits_{\nu_\mathrm{thr}(Q^2)}^{\nu_\mathrm{max}(Q^2)}\frac{\mathrm{d}\nu}{\nu} 
    \left[F_1(\nu,Q^2)-F_1(\nu,0)\right]\,.
\end{align}
In practice, we use
\begin{align}
    \nu_\mathrm{max}(Q^2) & = \nu_\mathrm{thr}(Q^2) + \Delta\nu  \nonumber\\
    & = \frac{Q^2+E_d^2}{2M_d}+E_d+\Delta\nu\,,
\end{align}
with $\Delta\nu$ set to $m_\pi$. Note that the same upper limit is used in the calculation of $\Delta E^\mathrm{inel}$; we thus separate the energy region corresponding to the internal structure of the nucleons in the inelastic contribution, following Ref.~\cite{Carlson:2013xea}.
The contribution of the hadron structure to $\Delta E^\mathrm{inel}$ can be calculated using empirical data on the deuteron (virtual) photoabsorption at energies starting from the pion production threshold, as also done in Ref.~\cite{Carlson:2013xea}. 
The sensitivity of both the subtraction and the inelastic contribution to the choice of $\Delta\nu$ is practically negligible in the interval $m_\pi/2\lesssim \Delta\nu\leq m_\pi$,
as discussed in more detail in Sec.~\ref{sec:results_TPE}.

\subsection{Longitudinal and transverse terms}
It is useful to compare the contributions to $\Delta E_{n\ell}$ with the conventional longitudinal and transverse contributions. The corresponding longitudinal and transverse VVCS amplitudes are related to the amplitudes $T_{1,2}(\nu,Q^2)$ through~\cite{Drechsel:2002ar}
\begin{align}
\label{eq:fLfT}
  f_T(\nu,Q^2) &= T_1(\nu, Q^2)\,, \\
\nonumber
 f_L(\nu, Q^2) &=-T_1(\nu,Q^2) +\left(1+\frac{\nu^2}{Q^2}\right)T_2(\nu,Q^2)\,,
\end{align}
which can be rearranged as 
\begin{align}
\label{eq:T12toTL}
    T_1(\nu,Q^2) &= f_T(\nu,Q^2) \,,\\
\nonumber
   T_2(\nu,Q^2)& = \frac{Q^2}{Q^2+\nu^2}\left[f_L(\nu,Q^2)+f_T(\nu,Q^2)\right]\,.
\end{align}
One can now clearly see that the contribution of $T_1$ is completely transverse, whereas that of $T_2$ has both a longitudinal and a transverse part. Note also that in order for these definitions to be consistent with the usual expectation that $f_L(\nu,Q^2)\propto Q^2$ at low $Q^2$, the Thomson term has to be subtracted from $T_1(\nu,Q^2)$. 

The imaginary parts of $f_{L,T}$ are related to the longitudinal and the transverse response functions by 
\begin{align}
    \im f_T(\nu,Q^2) & = \frac{1}{8}R_T(\nu,Q^2)\,,\\
    \im f_L(\nu,Q^2) & = \frac{1}{4}\frac{Q^2}{Q^2+\nu^2}R_L(\nu,Q^2)\,. 
\end{align}
It is evident that both $f_L$ and $f_T$ require a subtraction in the respective dispersive relations; apart from that, the TPE contribution to the Lamb shift can be rewritten in terms of the longitudinal and transverse components as done above. However, since the set of amplitudes $T_1$ and $T_2$ has better convergence properties, what we actually want is to decompose the sum of Eqs.~\eqref{eq:contrib_inelastic} and~\eqref{eq:contribution_subtraction_final} into the longitudinal and transverse contributions, keeping them written in terms of $F_1$ and $F_2$.
This is easily achieved using the relations between the two sets of VVCS amplitudes. Note that the $F_2$ term in Eq.~\eqref{eq:contrib_inelastic} only enters the longitudinal part of $\Delta E^\mathrm{inel}_{n\ell}$, whereas the $F_1$ terms in Eqs.~\eqref{eq:contrib_inelastic} and~\eqref{eq:contribution_subtraction_final} appear in both the transverse and longitudinal parts. The resulting formulas read:
\begin{widetext}
\begin{align} 
\label{eq:contrib_inelastic_L}
\Delta E_{n\ell,L}^\mathrm{inel}
&=-\frac{2\alpha^2}{M_d m}[\phi_{n\ell}(0)]^2
\int\limits_0^\infty\frac{\mathrm{d}Q^2}{Q^2} \!\!\!
\int\limits_{\nu_\mathrm{thr}(Q^2)}^\infty  \!\!\!
\frac{\mathrm{d}\nu}{\nu} 
\left[\tilde\gamma_L(\tau,\tau_l) F_1(\nu,Q^2)
+\frac{M_d\nu}{Q^2}\tilde\gamma_2(\tau,\tau_l)F_2(\nu,Q^2)\right]\,,\\
\label{eq:contrib_inelastic_T}
\Delta E_{n\ell,T}^\mathrm{inel}
&=-\frac{2\alpha^2}{M_d m}[\phi_{n\ell}(0)]^2
\int\limits_0^\infty\frac{\mathrm{d}Q^2}{Q^2} \!\!\!
\int\limits_{\nu_\mathrm{thr}(Q^2)}^\infty  \!\!\!
\frac{\mathrm{d}\nu}{\nu} \,\tilde\gamma_T(\tau,\tau_l) F_1(\nu,Q^2)\,,\\
\label{eq:contribution_subtraction_final_L}
    \Delta \bar{E}^\mathrm{subt}_{n\ell,L/T} & = \frac{8\alpha^2}{M_d}\left[\phi_{n\ell}(0)\right]^2
    \int\limits_0^\infty \frac{\mathrm{d}Q}{Q^2}\,\gamma_{L/T}(\tau_l)
    \int\limits_{\nu_\mathrm{thr}(Q^2)}^{\nu_\mathrm{max}(Q^2)}  \!\!\!
 \frac{\mathrm{d}\nu}{\nu} 
    \left[F_1(\nu,Q^2)-F_1(\nu,0)\right]\,,
%\label{eq:contribution_subtraction_final_T}
%    \Delta \bar{E}^\mathrm{subt}_{n\ell,T} & = \frac{8\alpha^2}{M_d}\left[\phi_{n\ell}(0)\right]^2
%    \int\limits_0^\infty \frac{\mathrm{d}Q}{Q^2}\,\gamma_T(\tau_l)
%    \int\limits_{\nu_\mathrm{thr}(Q^2)}^{\nu_\mathrm{max}(Q^2)}\frac{\mathrm{d}\nu}{\nu} 
%    \left[F_1(\nu,Q^2)-F_1(\nu,0)\right]
\end{align}
\end{widetext}
for the inelastic contributions and
 the subtraction contributions, respectively. The weighting functions that appear here are
\begin{align}
\gamma_L(x) & = \sqrt{1+x}-\sqrt{x}\,,\\
\gamma_T(x) & = 2x^{3/2}-2x\sqrt{1+x}+\sqrt{x}\,,\\
\tilde{\gamma}_{L,T}(x,y) & = \frac{\sqrt{x}\,\gamma_{L,T}(x)-\sqrt{y}\,\gamma_{L,T}(y)}{x-y}\,.
\end{align}
It is straightforward to see that $\gamma_{L}(x)+\gamma_{T}(x)=\gamma_1(x)$, and $\tilde{\gamma}_L(x,y)+\tilde{\gamma}_T(x,y)=\tilde{\gamma}_1(x,y)$. 

Note that this definition of the ``longitudinal" part only has a physical meaning for the sum $\Delta E_{n\ell,L}^\mathrm{inel}+\Delta \bar{E}_{n\ell,L}^\mathrm{subt}$, namely, the terms proportional to $F_1$ in this sum cancel the transverse part of the contribution of $F_2$, making the sum indeed a longitudinal quantity. The sum of the quantities $\Delta E^\mathrm{inel}_{n\ell,T}$ and $\Delta \bar{E}_{n\ell,T}$ (each being truly transverse), in turn, gives the transverse contribution to the non-pole part of $\Delta E_{n\ell}$.

\subsection{Comparison to Rosenfelder's method}
\label{sec:comparison}

Here we show that the covariant method used in the present work is (almost) equivalent to the formalism of Rosenfelder~\cite{Rosenfelder:1983aq}. We start from Eq.~\eqref{eq:lambshift_1} for the TPE contribution, assuming that the pole terms are subtracted from $T_1$ and $T_2$. We furthermore assume that the Thomson term is subtracted from $T_1$ both in $f_L$ and in $f_T$. As argued in Ref.~\cite{Birse:2012eb}, a more consistent way to subtract the amplitude would be to remove the Born term instead of just the pole plus Thomson term; the effect of this additional subtraction, however, would be very small [of the relative order of $\alpha r_d^2 / (6M\alpha_{E1}) \sim 10^{-3}$], so we do not consider the possible effect of doing the subtraction differently in the two methods.

Recalling that the definitions in Ref.~\cite{Rosenfelder:1983aq} differ from those in Eq.~\eqref{eq:fLfT} by a factor of $\mathbf{q}^2/Q^2$ for $f_L$ and $2$ for $f_T$,
\begin{equation}
f_L^{(\mathrm{R})}=\frac{\mathbf{q}^2}{Q^2}f_L,\quad f_T^{(\mathrm{R})}=2f_T\,,
\end{equation}
one gets, substituting the definitions into Eq.~\eqref{eq:lambshift_1}:
\begin{align}
\nonumber
\Delta E_{n\ell}&=-\frac{8\alpha^2 m}{i\pi}\left[\phi_{n\ell}(0)\right]^2 \int\dd^4 q
\frac{1}{(Q^4-4m^2\nu^2)}  \\
\label{eq:lambshift_R}
&
\quad\times\left\{\frac{1}{\mathbf{q}^2}f^{(\mathrm{R})}_L(\nu,Q^2) \right. +\\
\nonumber
&\qquad + \left. \frac{\nu^2}{Q^4}\left[f^{(\mathrm{R})}_T(\nu,Q^2)-f^{(\mathrm{R})}_T(0,0)\right]\right\}\,.
\end{align}
At this point, in order to arrive at Rosenfelder's formalism, one only needs to consider $f_{L,T}$ as functions of $(q_0,\mathbf{q}^2)$ rather than $(\nu,Q^2)$, assume that they obey unsubtracted DRs in $q_0$, substitute those DRs in Eq.~\eqref{eq:lambshift_R} (taking care of the subtracted $f_T^\mathrm{(R)}$ using the same DRs), and integrate over $q_0$. One has to note that our use of Eq.~\eqref{eq:nuclear_SR} is very similar to the use of an unsubtracted DR. Indeed, extending the upper integration limit in this equation to infinity, we get, upon combining it with the subtracted DR for $T_1$ given by Eq.~\eqref{eq:DRs},
\begin{eqnarray}
&T_1(\nu,Q^2)-T_1(0,0) 
=  \qquad\qquad\qquad\qquad\qquad\quad\\
\nonumber
&=
\frac{1}{2\pi M_d}\int\limits_{\nu_\mathrm{thr}(Q^2)}^\infty\mathrm{d}\nu'
\left\{\frac{\nu' F_1(\nu',Q^2)}{\nu'^2 -\nu^2-i0} 
- \frac{F_1(\nu',0)}{\nu'}\right\}\,.
\label{eq:DR_subtracted_inf}
\end{eqnarray}
Even though an unsubtracted DR in $q_0$ for functions of $(q_0,\mathbf{q}^2)$ cannot be directly obtained from this equation, the use of either $(\nu,Q^2)$ or $(q_0,\mathbf{q}^2)$ as independent variables should be mathematically equivalent as long as the analytic properties of the amplitudes are not spoiled by the choice of variables. Furthermore, taking the upper integration limit in Eq.~\eqref{eq:nuclear_SR} to infinity should also work fine if one uses the response functions calculated from a nuclear Hamiltonian, since this automatically filters out high-energy degrees of freedom, making the use of a finite $\nu_\mathrm{max}(Q^2)$ unnecessary. Under these conditions,  the two methods appear to be equivalent.

	The requirement that the analytic properties of the amplitudes should not be spoiled by the choice of variables is crucial for a reliable comparison of the two methods, and fulfilling it represents a certain difficulty in the perturbative scheme of $\chi$EFT, at least at lower orders in the expansion. Indeed, the definition of $T_2$ in terms of $f_T$ and $f_L$ in Eq.~\eqref{eq:T12toTL} introduces a pole at $\nu=\pm i Q$, where the denominator $Q^2+\nu^2=\mathbf{q}^2$ vanishes. This is the Siegert limit $|\mathbf{q}|\to 0$ at finite $\nu$, where the 0 in the denominator has to be compensated by the cancellation of the terms in the numerator~\cite{Drechsel:2002ar}, 
	\begin{equation}
	f_L(\nu,Q^2)+f_T(\nu,Q^2)=0 \text{ at } \nu = \pm iQ\,,
	\end{equation}
	which constrains the imaginary parts as well:
	\begin{equation}
	\lim_{\nu\to\pm iQ}\left[R_T(\nu,Q^2)+ 2\frac{Q^2}{Q^2+\nu^2}R_L(\nu,Q^2)\right]=0\,. 
	\end{equation}
	Current conservation will ensure that this constraint holds; however, the continuity equation relates the charge operator to current operators at higher orders in the $\chi$EFT expansion, i.e. current conservation is achieved only perturbatively. This, in turn, can lead to a violation of the equivalence of the two methods at a given $\chi$EFT order, especially at low orders in the expansion, complicating a meaningful comparison of their order-by-order results. This can be mitigated by the use of the Siegert theorem to calculate the transverse response. In practice, however, the difference between our Siegert and perturbative results at order $k^3$, as we see below, is already smaller than the estimated uncertainty due to higher-order terms in the expansion, indicating that the current conservation is achieved to an acceptable degree of precision.
	
	While the method of Rosenfelder does not seem to be sensitive to the Siegert limit due to the choice of independent variables ensuring that no pole appears in the dispersion relation, the associated singularity in the covariant method is simply a technical obstacle that makes comparison of the two methods somewhat difficult. The covariant method's use of the variables $(\nu, Q^2)$ is clearly more advantageous when a connection to empirical data is considered. 

The covariant method only uses the information from the physical domain $Q^2\ge 0$, thus allowing one to confront the calculated theoretical response functions with experimental data where the latter are available, and, ultimately, to replace the theory input with empirical information. In contrast to that, the use of $\mathbf{q}^2$ as the second independent variable requires information at all possible values of $\mathbf{q}^2$ for each $\nu\ge\nu_\mathrm{thr}$---also from the $Q^2<0$ region which is unphysical for scattering processes. While this circumstance does not preclude a correct theoretical calculation, connection to scattering experiments becomes more tenuous.

\section{Results for  muonic deuterium}
\label{sec:results_TPE}

\begin{figure}
\centering
    \includegraphics[width=\columnwidth]{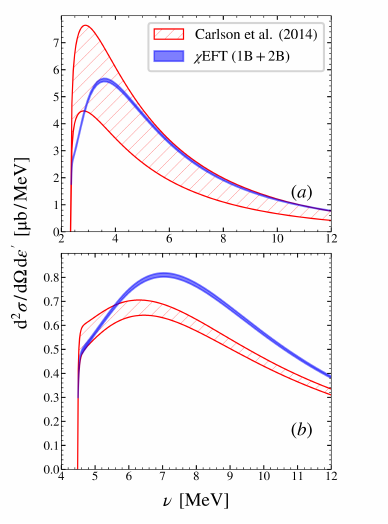}
    \caption{Electrodissociation cross section of the deuteron in kinematic regions where no data are available: comparison of the $\chi$EFT calculation (blue band) with the results in Ref.~\cite{Carlson:2013xea} with two different sets of fit parameters (hatched red band) at $(a)$
    $\epsilon=80~\mathrm{MeV},\,\theta=16^\circ$ and $(b)$ $\epsilon=180~\mathrm{MeV},\,\theta=30^\circ$.}
    \label{fig:electron_scatter_nodata}
\end{figure}

We now focus on the nuclear structure corrections to muonic deuterium, which constitute the main results of this paper.
Because our idea is to use response functions calculated from $\chi$EFT to inform the DR analysis,  we first
 compare $\chi$EFT results with the 
results obtained in Ref.~\cite{Carlson:2013xea}
for the kinematics where no data are available.
Figure~\ref{fig:electron_scatter_nodata} shows  a  selection of two different kinematics  where  the $\chi$EFT calculations are shown as blue bands, while results from Ref.~\cite{Carlson:2013xea} are shown as hatched red bands. The latter were obtained by extrapolating the fit of the available quasi-elastic electron scattering data to lower values of the electron beam energy and more forward angles.
Note that the difference between the lower and the upper red curves was used in Ref.~\cite{Carlson:2013xea} to estimate the uncertainty of the Lamb shift calculation. 
While the theory curves are in a reasonable agreement for the kinematics shown in Fig.~\ref{fig:electron_scatter_nodata}$(a)$, there are visible differences in Fig.~\ref{fig:electron_scatter_nodata}$(b)$ that can be attributed to the DR extrapolation. Most importantly,  the $\chi$EFT calculations have a  smaller uncertainty and, thus, supersede the existing DR calculations in terms of precision.
 It is to be expected that this will reflect in a reduced uncertainty in the TPE calculation provided by our  hybrid method.

At this point, by feeding the response functions computed within $\chi$EFT 
to the DR method we are ready to compute the various terms in Eqs.~(\ref{eq:contrib_inelastic_L}), (\ref{eq:contrib_inelastic_T}) and (\ref{eq:contribution_subtraction_final_L}) and analyze the various sources of uncertainties.

First, we analyze the uncertainty coming from the different parametrizations of the single-nucleon form factor.
 On one hand, we use the dipole model in which the electric form factors of the proton and the neutron are  respectively given by
\begin{equation}
 G_E^p(Q^2) = \frac{1}{\left(1+\frac{Q^2}{\Lambda_D^2}\right)^2}\,,
\end{equation}
and 
\begin{equation}
 G_E^n(Q^2) = -\mu_n \frac{Q^2}{4(m_n^2+Q^2)}\frac{1}{\left(1+\frac{Q^2}{\Lambda_D^2}\right)^2}\,,
\end{equation}
with the value $\Lambda_D=833$~MeV determined from fits to elastic electron scattering off the proton and deuteron~\cite{HydeWright:2004gh}, $\mu_n =-1.913~\mu_N$~\cite{PDG2020} is the neutron magnetic moment and $m_n$ is the neutron mass. The magnetic form factors are given by  $G_M^p = \mu_p/\left(1+Q^2/\Lambda_D^2\right)^2$ and $G_M^n = \mu_n/\left(1+Q^2/\Lambda_D^2\right)^2$. This parametrization is common in the  nuclear physics literature and will facilitate comparison with Refs.~\cite{Hernandez:2014pwa,Hernandez:2017mof,Hernandez:2019zcm}. 
On the other hand, we perform a model-independent expansion of the form factors up to terms linear in $Q^2$,
\begin{align}
\label{eq:linear_ff}
\nonumber
G^{p,n}_{E,M}(Q^2) &= G^{p,n}_{E,M}(0)\left[1-\langle {r^{p,n}_{E,M}}^2\rangle\,\frac{Q^2}{6}\right]\\
&\quad +\mathcal{O}(Q^4)\,,
\end{align}
where $G^{p,n}_{E,M}(0)$ and $\langle {r^{p,n}_{E,M}}^2\rangle$ are, respectively, the charge/magnetic moments and mean-square charge/magnetization radii.
Finally, we also implement the realistic form factor from Refs.~\cite{Kelly:2004hm} and the $z$-expansion-based form factors from Ref.~\cite{Tomalak}.

Table~\ref{tab:formfactors} shows the different contributions to the energy shift of the $2S$ state in the impulse approximation, i.e., with only 1B currents, for several nucleon form factor parametrizations. 
Compared to the point particle limit, the magnitude of the total nuclear structure correction decreases by about 2\% when we use nucleon form factors. We use two variants of the linear form factors of Eq.~\eqref{eq:linear_ff}. The ``PDG 2020" results use the most recent recommendations for charge and magnetization radii from the Particle Data Group~\cite{PDG2020}, including the muonic hydrogen radius for the proton. We also present the ``CODATA 2014" results, which use linear form factors with CODATA 2014 averages~\cite{Mohr:2015ccw} because the dipole form factors are also fit to older $ep$ measurements~\cite{HydeWright:2004gh} which are inconsistent with the PDG 2020 value for the proton charge radius. We see in Table~\ref{tab:formfactors}, however, that the differences among the two linear form factors, as well as the differences among the Kelly, dipole, and $z$-expansion form factors, are much smaller than 1\% and, as we see further below, are also much smaller than the nuclear structure uncertainty due to the $\chi$EFT expansion. All of the following results are obtained with the recent parametrization of form factors based on $z$-expansion~\cite{Tomalak} unless explicitly stated otherwise. This form factor is optimal for our analysis as it is model independent, incorporates the low-$Q^2$ constraint from the muonic-hydrogen spectroscopy of the proton radius, and is optimized for  $Q^2\lesssim$~a~few~GeV$^2$.

\begin{widetext}
\begin{center}
\begin{table*}[htbp]
\begin{tabularx}{\columnwidth}{lYYYYYY}
                                       & 1B, point particle  & 1B, linear\qquad\qquad(PDG 2020) & 1B, linear (CODATA 2014) & 1B, Kelly & 1B, dipole 
                                       & 1B, $z$-expansion\\

\hline
\hline
$\Delta E_{2S}^\mathrm{inel}$ [meV]    &                &       &    &     &  &   \\
%$\qquad$ --- due to $F_1$   & $-0.298$ & $-0.270$ & $-0.270$ &  $-0.272$ &  $-0.274$  &    \\
%$\qquad$ --- due to $F_2$   & $-1.627$ & $-1.584$ & $-1.582$ &  $-1.586$ &  $-1.589$ &  \\
$\qquad$ --- ``transverse"     \rule{0pt}{3.5ex}
                            & $-0.077$ & $-0.068$ & $-0.068$ &  $-0.069$ &  $-0.070$ & $-0.069$ \\
$\qquad$ --- ``longitudinal"   
                            & $-1.848$ & $-1.786$ & $-1.783$ &  $-1.789$ &  $-1.793$ &  $-1.791$\\

total    \rule{0pt}{3.5ex}
                            & $-1.925$ & $-1.854$ & $-1.851$ &  $-1.858$ &  $-1.863$ &  $-1.861$\\
\hline
\hline
$\Delta \bar{E}_{2S}^\mathrm{subt}$ [meV]            &     &    &  &  & &  \\ 
$\qquad$
--- ``transverse"  &$\hM 0.075$ &$\hM 0.065$ &$\hM 0.065$ &$\hM 0.066$ &$\hM 0.067$ & $\hM 0.066$ \\
$\qquad$
--- ``longitudinal" &$\hM 0.354$ &$\hM 0.328$ &$\hM 0.328$ &$\hM 0.330$ &$\hM 0.332$ &  $\hM 0.331$\\
total     \rule{0pt}{3.5ex}
                 &$\hM 0.429$ &$\hM 0.393$ &$\hM 0.393$ &$\hM 0.396$ &$\hM 0.399$ & $\hM 0.398$ \\
\hline
\hline 
$\Delta E_{2S}^\mathrm{inel}+\Delta \bar{E}_{2S}^\mathrm{subt}$  [meV]      &    & &    &    & \\
$\qquad$ --- transverse   & $-0.002$ & $-0.003$ & $-0.003$ &  $-0.003$ &  $-0.003$ &  $-0.003$\\
$\qquad$ --- longitudinal  & $-1.493$ & $-1.458$ & $-1.455$ &  $-1.459$ &  $-1.461$ & $-1.460$ \\
total  \rule{0pt}{3.5ex}
                           & $-1.495$ & $-1.461$ & $-1.458$ &  $-1.462$ &  $-1.464$ & $-1.463$ \\
\end{tabularx}
\caption{Impulse-approximation results for the nuclear structure contributions to the energy shift of the $2S$ state for various nucleon form factor parametrizations: the point particle limit, linear in $Q^2$ with the PDG 2020~\cite{PDG2020} recommendations for all the mean-square charge and magnetization radii including the value of 0.84087~fm for the proton charge root-mean-square (rms) radius, the same but using the CODATA 2014~\cite{Mohr:2015ccw} recommendation of 0.8751~fm for the proton charge rms radius value, the Kelly, traditional dipole forms, and recent extraction based on $z$ expansion form factors, as explained in the text. 
\label{tab:formfactors}
}
\end{table*}
\end{center}
\end{widetext}

In Table~\ref{tab:current_conservation}, we present the contributions to the energy shift of the $2S$ state in impulse approximation as well as with the meson-exchange current contributions included in two ways: using the ``Siegert" method which assumes current conservation as discussed above and explicit calculation of the matrix elements with 1B and 2B currents. The meson-exchange contribution is about 2.5\% of the total nuclear structure correction. 

\begin{widetext}

%%%Dipole form-factor results:
%\begin{center}
%\begin{table*}%[htb]
%\begin{tabularx}{\columnwidth}{lYYY}
%                                         & 1B & Siegert  & Full = 1B+2B \\

%\hline
%\hline
%$\Delta E_{2S}^\mathrm{inel}$ [meV]     &         &        &        \\
%$\qquad$ --- due to $F_1$     &  $-0.274$   & $-0.260$  & $-0.296$ \\
%$\qquad$ --- due to $F_2$    &  $-1.589$ & $-1.631$     & $-1.638$\\
%$\qquad$ --- transverse     \rule{0pt}{3.5ex}
%                             &  $-0.070$  & $-0.065$    & $-0.075$\\
%$\qquad$ --- longitudinal   
%                              &  $-1.793$ & $-1.825$    & $-1.858$\\

%total    \rule{0pt}{3.5ex}
%                              &  $-1.863$ & $-1.890$    & $-1.934$\\
%\hline
%\hline
%$\Delta \bar{E}_{2S}^\mathrm{subt}$ [meV]     &    &        &\\ 
%$\qquad$
%--- transverse   &$\hM 0.067$ &$\hM 0.059$   & $\hM 0.070$\\
%$\qquad$
%--- longitudinal &$\hM 0.332$ &$\hM 0.312$   & $\hM 0.351$\\
%total     \rule{0pt}{3.5ex}
%                 &$\hM 0.399$ &$\hM 0.371$   & $\hM 0.422$ \\
%\hline
%\hline 
%$\Delta E_{2S}^\mathrm{inel}+\Delta \bar{E}_{2S}^\mathrm{subt}$  [meV] &    &   & \\
%$\qquad$ --- transverse    &  $-0.003$ & $-0.006$  & $-0.005$\\
%$\qquad$ --- longitudinal   &  $-1.461$ &$-1.514$  & $-1.507$\\
%total  \rule{0pt}{3.5ex}
%                            &  $-1.464$ & $-1.520$  & $-1.512$\\
%%%\end{tabularx}
%%%End dipole form factor results
\begin{center}
\begin{table*}%[htb]
\begin{tabularx}{\columnwidth}{lYYY}
                                         & 1B & Siegert  & Full = 1B+2B \\

\hline
\hline
$\Delta E_{2S}^\mathrm{inel}$ [meV]     &         &        &        \\
%$\qquad$ --- due to $F_1$     &  $-0.274$   & $-0.260$  & $-0.296$ \\
%$\qquad$ --- due to $F_2$    &  $-1.589$ & $-1.631$     & $-1.638$\\
$\qquad$ --- ``transverse"     \rule{0pt}{3.5ex}
                             &  $-0.069$  & $-0.065$    & $-0.075$\\
$\qquad$ --- ``longitudinal"   
                              &  $-1.791$ & $-1.824$    & $-1.857$\\

total    \rule{0pt}{3.5ex}
                              &  $-1.861$ & $-1.889$    & $-1.932$\\
\hline
\hline
$\Delta \bar{E}_{2S}^\mathrm{subt}$ [meV]     &    &        &\\ 
$\qquad$
--- ``transverse"   &$\hM 0.066$ &$\hM 0.059$   & $\hM 0.070$\\
$\qquad$
--- ``longitudinal" &$\hM 0.331$ &$\hM 0.311$   & $\hM 0.351$\\
total     \rule{0pt}{3.5ex}
                 &$\hM 0.398$ &$\hM 0.370$   & $\hM 0.421$ \\
\hline
\hline 
$\Delta E_{2S}^\mathrm{inel}+\Delta \bar{E}_{2S}^\mathrm{subt}$  [meV] &    &   & \\
$\qquad$ --- transverse    &  $-0.003$ & $-0.006$  & $-0.005$\\
$\qquad$ --- longitudinal   &  $-1.460$ &$-1.513$  & $-1.506$\\
total  \rule{0pt}{3.5ex}
                            &  $-1.463$ & $-1.519$  & $-1.511$\\
\end{tabularx}
\caption{Results for the nuclear structure contributions to the energy shift of the $2S$ state in impulse approximation (1B), 
with the current operators deduced from the charge operator assuming current conservation (Siegert) and with both one- and two-body currents explicitly included (Full). 
\label{tab:current_conservation}
}
\end{table*}
\end{center}
\end{widetext}

We present order-by-order results in the $\chi$EFT expansion in Table~\ref{tab:orders}. 
The  error estimate is again calculated as:
\begin{align}
\sigma(\Delta E) %& = \max\left\{
%\chi^4 \left|\Delta E^{(k^0)}\right|,\ \chi^3 \left|\Delta E^{(k^1)}-\Delta E^{(k^0)}\right|,\ \chi^2 \left|\Delta E^{(k^2)}-\Delta E^{(k^1)}\right|,\ \chi \left|\Delta E^{(k^3)}-\Delta E^{(k^2)}\right|\right\}\\
& = \max\left\{0.007, 0.006, 0.007, 0.012 \right\}~\mathrm{meV}\nonumber \\
& = 0.012~\mathrm{meV}\,.
\label{eq:cheft_uncertainty}
\end{align}
 To arrive at this estimate, we assumed that the typical momentum scale that contributes the most to the response functions is $k\simeq m_\pi$, and we take $\Lambda_b=500$~MeV equal to the cutoff used in the NN interaction. The values of the error estimate above are given for the total (inelastic$+$subtraction) contribution, resulting in
\begin{equation}
\Delta E^\mathrm{inel+subt}_{2S} = -1.511(12)~\mathrm{meV}\,,
\label{eq:inel_plus_subt}
\end{equation}
with the relative uncertainty of $0.8\%$. The difference between the $k^{3+}$ and the $k^3$ results, generated by the $k^4$ terms in the NN interaction, is well within this uncertainty  estimate.

The sensitivity of the subtraction (and inelastic) contribution to the choice of $\nu_\mathrm{max}(Q^2)$ can be estimated by varying this limit. We find that using $\nu_\mathrm{max}(Q^2)=\nu_\mathrm{thr}(Q^2)+100~\text{MeV}$ changes the total result for $\Delta  \bar{E}^\mathrm{subt}_{2S}$ and $\Delta  E^\mathrm{inel}_{2S}$ by less than $0.003$~meV. The sensitivity to the choice of
the upper limit of the $\nu$ integration thus appears to be much smaller than the uncertainty due to higher orders in the $\chi$EFT expansion~(\ref{eq:cheft_uncertainty}), and contributes a negligible amount to the total uncertainty estimate. To check this, we used a more extreme
cut-off, setting $\nu_\mathrm{max}(Q^2)=\nu_\mathrm{thr}(Q^2)+m_\pi/2$; even that changes the values of $\Delta  \bar{E}^\mathrm{subt}_{2S}$ and $\Delta  E^\mathrm{inel}_{2S}$ by only $0.002$ and $0.005$ meV, respectively (although the small transverse contributions experience the relatively significant change of about $0.6-0.9\%$).

To obtain the value of the nuclear structure contribution to the TPE correction,
we add to the result of Eq.~(\ref{eq:inel_plus_subt}) the elastic part of the energy shift, which, for the $2S$ level, is evaluated in Ref.~\cite{Carlson:2013xea} at
\begin{align}
\Delta \bar{E}^\mathrm{el}_{2S}=-0.417(2)\text{ meV}\,,
\label{eq:elastic2S}
\end{align}
where the uncertainty is estimated by taking different form factor parametrizations
deduced in~\cite{Abbott:2000ak}. As explained above, we also have to add the hadronic contribution, coming from energy scales typical of the pion production and higher. We use the value calculated in~\cite{Carlson:2013xea}:
\begin{align}
\Delta E^\mathrm{hadr}_{2S}=-0.028(2)\text{ meV}\,.
\label{eq:hadronic}
\end{align}
The Coulomb distortion contribution, which is a nominally subleading---$O(\alpha^6\ln\alpha)$---but
practically important effect, has to be taken into account as well. We use the value
from the compilation in Ref.~\cite{Krauth:2015nja}:
\begin{align}
\Delta E^\mathrm{Coulomb}_{2S}=0.262(2)\text{ meV}\,.
\label{eq:Coulomb}
\end{align}
Adding these contributions together, we get the total two-photon-exchange contribution to the $\mu$D
Lamb shift:
\begin{align}
    \Delta E^\mathrm{TPE}_{2S} = -1.695(13)\text{ meV}\,.
\end{align}

\begin{widetext}
\begin{center}
\begin{table*}[h]
\begin{tabularx}{\columnwidth}{lYYYYYY}
                                                & $k^0$        & $k^1$        & $k^2$        & $k^3$        & Full $=k^{3+}$ \\
\hline
\hline
$\Delta E_{2S}^\mathrm{inel}$ [meV]             &              &              &              &              &                                \\
%$\qquad$ --- due to $F_1$                       &$\hM 0   $    &$-0.279$    & $-0.277$   &  $-0.295$  & $-0.296$    \\
%$\qquad$ --- due to $F_2$                       &$-1.123$    &$-1.502$    & $-1.590$   &  $-1.634$  & $-1.638$    \\
$\qquad$ --- ``transverse"     \rule{0pt}{3.5ex}                    
                                                & $\hM 0   $    &$-0.071$    & $-0.070$   &  $-0.075$  & $-0.075$    \\
$\qquad$ --- ``longitudinal"   
                                                &$-1.122$    &$-1.709$    & $-1.795$   &  $-1.851$  & $-1.857$    \\
total    \rule{0pt}{3.5ex} 
                                                &$-1.122$    &$-1.780$    & $-1.865$   &  $-1.926$  & $-1.932$    \\
\hline
\hline
$\Delta \bar{E}_{2S}^\mathrm{subt}$ [meV]             &              &              &              &              &               \\ 
$\qquad$
--- ``transverse"                                  &$\hM 0      $ &$\hM 0.067$ &$\hM 0.067$ &$\hM 0.070$ &$\hM 0.070$  \\
$\qquad$
--- ``longitudinal"                                &$\hM 0      $ &$\hM 0.337$ &$\hM 0.335$ &$\hM 0.349$ &$\hM 0.351$  \\
total     \rule{0pt}{3.5ex}
                                                &$\hM 0      $ &$\hM 0.404$ &$\hM 0.402$ &$\hM 0.419$ &$\hM 0.421$  \\
\hline
\hline 
$\Delta E_{2S}^\mathrm{inel}+\Delta \bar{E}_{2S}^\mathrm{subt}$  [meV] &    &    &        &     \\
$\qquad$ --- transverse                      &$\hM 0   $     &$-0.003$     &$-0.003$   &  $-0.005$ & $-0.005$    \\
$\qquad$ --- longitudinal                     &$-1.122$     &$-1.373$     &$-1.460$   &  $-1.502$  & $-1.506$    \\
total  \rule{0pt}{3.5ex}
                                               &$-1.122$     &$-1.376$     &$-1.463$   &  $-1.507$  & $-1.511$    \\
\end{tabularx}
\caption{Nuclear structure contributions to the energy shift of the $2S$ state for various $\chi$EFT orders. The ``full'' ($k^{3+}$) result includes $k^{4}$ terms in the interactions, but not in the currents. 
\label{tab:orders}
}
\end{table*}
\end{center}
\end{widetext}

\section{Conclusion}

We have performed an analysis of the nuclear polarizability correction to the Lamb shift of muonic deuterium by combining dispersion relations with chiral effective field theory. This approach relies only on controlled approximations and provides a well-justified uncertainty estimate. Final results for the two-photon-exchange correction to the Lamb shift are compared to those of prior $\chi$EFT~\cite{Hernandez:2014pwa,Hernandez:2017mof,Hernandez:2019zcm} and dispersive~\cite{Carlson:2013xea} studies in Table~\ref{tab:comparison}.

\begin{center}
\begin{table}%[htb]
\begin{tabularx}{0.67\columnwidth}{ll}
                                       & $\Delta E^\mathrm{TPE}_{2S}$ [meV]  \\
                                        
\hline
\hline
This work    &                  \\
$\qquad$ --- 1B+2B\quad    \rule{0pt}{3.5ex} &  -1.695(13)\\
$\qquad$ --- Siegert\quad  &  -1.703(15)\\
Ref.~\cite{Pachucki:2011xr}      \rule{0pt}{3.5ex}  &  -1.680(16)  \\
Ref.~\cite{Pachucki:2015uga}     \rule{0pt}{3.5ex}  &  -1.717(20)  \\
Ref.~\cite{Hernandez:2014pwa}    \rule{0pt}{3.5ex}  &  -1.690(20)  \\
Ref.~\cite{Hernandez:2017mof}    \rule{0pt}{3.5ex}  &  -1.712(21)  \\
Ref.~\cite{Hernandez:2019zcm}    \rule{0pt}{3.5ex}  &  -1.703      \\
Ref.~\cite{Carlson:2013xea}      \rule{0pt}{3.5ex}  &  -2.011(740) \\
\end{tabularx}
\caption{ Comparison of our dispersive $\chi$EFT result for the two-photon exchange corrections to the $\mu D$ Lamb shift with prior calculations. For comparison with Ref.~\cite{Hernandez:2017mof}, we use the value obtained with the same $\chi$EFT interactions. Their full result, obtained by averaging the values given by different $\chi$EFT interactions, is $-1.715^{+22}_{-24}$~meV. For consistent comparison with Ref.~\cite{Hernandez:2019zcm}, we have applied the hadronic correction of Eq.~\eqref{eq:hadronic} to their ``$\eta$-less'' result. 
\label{tab:comparison}
}
\end{table}
\end{center}

We obtain good agreement with prior studies as shown in Table~\ref{tab:comparison}. In Refs.~\cite{Pachucki:2011xr, Pachucki:2015uga,  Hernandez:2014pwa,Hernandez:2017mof,Hernandez:2019zcm}, the polarizability correction to the Lamb shift was directly evaluated without explicitly calculating the electromagnetic response functions. The calculations in Refs.~\cite{Pachucki:2011xr, Pachucki:2015uga, Hernandez:2014pwa,Hernandez:2017mof} were performed in a perturbative framework using the AV18~\cite{Pachucki:2011xr, Pachucki:2015uga, Hernandez:2014pwa} and $\chi$EFT~\cite{Hernandez:2014pwa,Hernandez:2017mof} interactions. 
Reference~\cite{Hernandez:2019zcm} was based on the Rosenfelder~\cite{Rosenfelder:1983aq} formalism, which is, in principle, equivalent to our method based on dispersion relations as discussed in Sec.~\ref{sec:comparison}. We reiterate, however, that our approach allows us to more directly compare the inputs that go into the dispersion relations to a large body of existing data from electron-scattering experiments. Indeed, an excellent agreement with electron-scattering data is obtained over a wide range of kinematics. Without sacrificing model independence, we achieve a much higher precision than the fully data-driven approach of Ref.~\cite{Carlson:2013xea} which used extrapolants fitted to electron-scattering data to inform their dispersion relations. This was possible because $\chi$EFT, which is based on symmetries of quantum chromodynamics and is constrained by $\pi$N and NN data, allows us to obtain accurate and precise electromagnetic response functions even in kinematics not explored so far by electrodissociation experiments. While $\chi$EFT served as a good proxy for electron-scattering experiments here, even at low beam energies ($\lesssim 200$~MeV) and forward angles ($\lesssim 30^\circ$), future measurements of the deuteron electrodisintegration at the MAMI A1 and MESA facilities will enable a more comprehensive comparison of the $\chi$EFT cross sections with experiments and allow us to fully exploit the merits of the dispersive approach.

\section*{Acknowledgements}
This work was supported by the Deutsche Forschungsgemeinschaft (DFG, German Research Foundation), 
in part through the Collaborative Research Center [The Low-Energy Frontier of the Standard Model, Projektnummer 204404729 - SFB 1044], 
and in part through the 
Cluster of Excellence [Precision Physics, Fundamental Interactions, and Structure of Matter] (PRISMA$^+$ EXC 2118/1) 
within the German Excellence Strategy (Project ID 39083149). 
This work was supported by EU Horizon 2020 research and innovation programme, STRONG-2020 project 
under grant agreement No 824093, and by the German-Mexican research collaboration Grant No. 278017 (CONACyT) and No. SP 778/4-1 (DFG).

\end{document}